\documentclass{epl}

\title{
Rate- and state-dependent friction law \\
and statistical properties of earthquakes
}
\author{Akio Ohmura and Hikaru Kawamura}

\institute{                    
Department of Earth and Space Science, Faculty of Science,
Osaka University, \\
Toyonaka 560-0043, Japan
}
\pacs{nn.mm.xx}{91.30.Ab}
\begin{document}

\maketitle

\begin{abstract}
In order to clarify how the statistical properties of earthquakes depend on the
constitutive law characterizing the stick-slip dynamics, we make an extensive
numerical simulation of the one-dimensional
spring-block model with a rate- and state-dependent friction law.
Both the magnitude distribution and  
the recurrence-time distribution are studied with varying the
constitutive parameters characterizing the model. While a 
continuous spectrum of seismic events
from smaller to larger magnitudes is obtained, earthquakes described 
by this model turn out to possess pronounced ``characteristic'' 
features.
\end{abstract}

%\section{Section title}
Recent studies have revealed that an earthquake can be regarded as a stick-slip frictional instability which a natural fault
driven by steady motions of tectonic plates exhibits. Hence, an earthquake
occurrence is governed by
the physical law of rock friction \cite{Scholzbook,ScholzRev}.
Unfortunately, our present understanding of physics of friction is still poor.
We do not have precise knowledge of the constitutive law characterizing
the stick-slip dynamics of earthquake faults.
The difficulty lies partly in 
the fact that a complete microscopic theory of 
friction is still not available, but also in the fact that the length and
time scales relevant to earthquakes are so large that the applicability
of laboratory experiments on rock friction is not necessarily clear.

Detailed characteristics of the friction force are specified 
by the constitutive relation \cite{Scholzbook,ScholzRev,Dieterich,Ruina}.
One fundamental question in earthquake studies
might be how the properties of earthquakes depend on the constitutive
law and other material parameters 
characterizing earthquake faults.
To answer this question and to get deeper insight into the physical 
mechanism governing the stick-slip process of earthquakes, 
proper modeling of an earthquake is essential. 
Indeed, earthquake models of various
levels of simplifications  have been proposed in geophysics
and statistical physics, and their statical properties have been extensively
studied mainly by means of numerical computer simulations.
In model simulations, one can easily control the constitutive parameters,
which is almost impossible for natural faults.

One of the standard model used in statistical studies of
earthquakes is the so-called 
spring-block model originally proposed by Burridge and Knopoff 
(BK model) \cite{BK}. 
The BK model, combined
with several types of constitutive relations, have been extensively studied
by numerical simulations 
\cite{CL,Carlson91,Carlson94,Shaw,Myers,MoriKawamura,Kawamura,Xia}. 
In order for the model to exhibit a 
dynamical instability corresponding to an earthquake, it is essential 
that the friction force exhibits a frictional {\it weakening\/}
property, {\it i.e.\/},
the friction should become weaker as the block slides.
One of the simplest form of the friction
force widely used 
is a velocity-weakening friction force 
\cite{Carlson91,Carlson94}. 
Here, the friction force is assumed to 
be a single-valued function 
of the velocity, getting smaller as the velocity increases.
The other form employed in the previous analyses
might be a slip-weakening friction force where
the friction force is assumed to be a single-valued function of 
the cumulative fault slip, 
getting smaller as the
slip distance increases \cite{Ida,Shaw,Myers}.

From laboratory
experiments of rock friction, however, there is an indication
that the real constitutive relation might be more complex,
neither purely velocity-weakening nor slip-weakening.
%For example, it has been observed that the static friction 
%increases gradually, logarithmically in time \cite{Dieterich}. 
%Since in this case the system is at rest (zero velocity) 
%staying at the same position 
%(zero slip distance), such an experimental observation 
%can be explained neither
%by the velocity-weakening constitutive law nor by the slip-weakening
%constitutive law. Rather, the experimental result 
Experiments suggest that the friction force depends
on some ``hidden'' variable, possibly representing the state
of the rock interface.
Some time ago,
Dieterich and Ruina proposed an empirical form of the constitutive
law, a rate- and state-dependent friction law, by
phenomenologically introducing the time-dependent
``state variable'' and its time-evolution equation 
\cite{Dieterich,Ruina}.
Though introduced empirically,
this rate- and state-dependent constitutive law is devised so
as to reproduce certain noticeable features of rock experiments
mentioned above. Tse and Rice employed 
this rate- and state-dependent constitutive relation 
in their numerical simulations of earthquakes \cite{TseRice}.
These authors studied
the stick-slip motion of the two-dimensional strike-slip fault
within an elastic continuum theory,
assuming that the fault motion is rigid along strike.
It was then observed that large events repeated quasi-periodically.
Since then, similar rate- and state-dependent constitutive laws 
have widely been used in
numerical simulations \cite{Stuart,Horowitz,Rice,Ben-ZionRice,
%\textit{Koto and Hirasawa}, 1999; 
Kato,Bizzarri}. Somewhat different type of slip- and state-dependent
constitutive law was also used  \cite{Cochard}.

Cao and Aki performed a numerical simulation of earthquakes
by combining the one-dimensional BK model with the rate- and state-dependent 
friction law in which various constitutive parameters were set nonuniform
over blocks \cite{CaoAki}. 
In the present letter, we wish to extend an earlier calculation by 
Cao and Aki to study the
{\it statistical properties\/} of the 1D BK model combined with the
rate- and state-dependent constitutive law 
with uniform constitutive parameters.
Such a calculation might be useful due to the following two reasons:
First, due to the simplicity of the
1D BK model, it is now possible to generate large number of
events for large enough system
to obtain various statistical properties of sufficient accuracy.
Second, by comparing the obtained results with the previous ones for
the 1D BK model with the purely velocity-weakening or slip-weakening
constitutive law with uniform constitutive parameters, 
it is possible to clarify the dependence of earthquake properties
on the underlying constitutive law.

%An obvious limitation of the BK model 
%is that an artificial discreteness 
%is introduced into the model
%associated with its intrinsic block structure. In particular, 
%Rice and collaborators argued that the slip complexity of the BK model
%might be caused by its intrinsic discreteness 
%\cite{Rice,Ben-ZionRice}.
%Although it is 
%desirable to elucidate the effect of discreteness by studying the
%continuum limit, it is also important to elucidate the statistical 
%properties of 
%the type of the model where the discrete BK structure is
%combined with the rate- and state-dependent friction law, to compare its
%statistical properties with those of the standard
%BK model with the
%velocity-weakening or slip-weakening friction law
%studied extensively in the past.

In the one-dimensional BK model, an earthquake fault is
simulated by an array of blocks, each of which is connected to the 
neighboring blocks via 
the elastic springs of the spring constant $k_c$, 
and to 
the moving plate via the springs of the spring constant $k_p$. 
All blocks are subject to the 
friction force $\phi$, the source of the nonlinearity in the model.

The equation of motion 
for the $i$-th block can be written as
\begin{eqnarray}
m\ddot{u'}_i=k_p(\nu 't'-u'_i)+k_c(u'_{i+1}-2u'_i+u'_{i-1}) 
- \phi (\dot{u'}_i, \theta'_i),
\end{eqnarray}
where $t'$ is the time, $u'_i$ the 
displacement of the 
$i$-th block, $m$ the mass of the block, and
$\nu'$ is the loading rate 
representing the speed of the plate.
The form of the friction force $\phi$ is specified by the constitutive 
relation, which, as mentioned above, is a vitally important, yet largely
ambiguous part in the proper description of earthquakes.
Here, as the form of the friction force,
we assume a rate- and state-dependent friction force, 
\begin{eqnarray}
\phi_i=\{c'
 +a'\log(1+\frac{v'_i}{v^*})
 +b'\log\frac{v^* \theta'_i}{{\cal L}}\}{\cal N}, 
\end{eqnarray}
where $v'_i=\dot{u'}_i$ is the velocity of the $i$-th block, 
$\theta' _i(t)$ is the time-dependent state variable (with the dimension of
time) representing the ``state'' of the slip interface, $v^*$ is
a reference velocity, ${\cal N}$ is an 
effective normal load, ${\cal L}$ is a characteristic slip distance
which is a measure of the distance of sliding necessary for the
surface to evolve to a new state,
while $a',\ b'$ and $c'$ are
numerical constants describing the rate- and state-dependent
friction law. 
The first term ($c$-term) is a constant taking a value around $\frac{1}{3}$,
which dominates
the total friction in magnitude \cite{Scholzbook}. The second term ($a$-term)
is a velocity-strengthening direct term describing 
the part of the 
friction which follows the velocity-change immediately. Note that we put a 
factor unity in the logarithm of this second term, which enables
one to describe the system at a complete halt, whereas, without this term,
the system cannot stop because of the logarithmic anomaly occurring at $v'=0$.
The third part ($b$-term) is an indirect term
dependent of the state variable and follows the velocity-change
via the time dependence of the state variable.
Laboratory experiments suggest that
the $a$- and $b$-terms are smaller than the $c$-term
by one or two orders of magnitudes \cite{Scholzbook,ScholzRev}, 
yet they play an essential role in stick-slip dynamics. 

There are several proposals for the form of 
the time-evolution equation of the state variable.
Here, we assume the so-called slowness law given by
\begin{eqnarray}
\frac{d\theta'_i}{dt'}=1-\frac{v'_i\theta'_i}{{\cal L}}.
\label{slowness}
\end{eqnarray}
%
%
%At complete stop, {\it i.e.\/}, $v'_i=0$ for all $i$, this evolution equation
%can easily be solved to give $\theta'_i=t'+c_i$ ($c_i$
%is a constant). Substituting this into Eq.() yields the static friction
%consistent with the observed logarithmic time dependence.
%When the block slips with high velocity, on the other hand,
%the r.h.s. of Eq.() becomes negative 
%and the state variable $\theta'_i$ drops rapidly with time. This
%eventually leads to the frictional weakening via the $b$-term in Eq.(), 
%leading to an earthquake instability.  

We take the length unit to be the characteristic slip distance ${\cal L}$,
and the time unit to be the rise time of an earthquake
$\omega^{-1}=(m/k_p)^{1/2}$, and set $v^* = {\cal L}\omega$.
The coupled equations of motion are then made dimensionless by introducing the 
dimensionless variables, $t=\omega t'$,
$u_i=u'_i/{\cal L}$,  
$v_i=v'_i/v^*$, 
$\theta_i=\theta'_i v^*/{\cal L}$, 
$\nu=\nu'/v^*$, 
$a=a'{\cal N}/(k_p{\cal L})$,
$b=b'{\cal N}/(k_p{\cal L})$,
and $c=c'{\cal N}/(k_p{\cal L})$,  
\begin{eqnarray}
\frac{d^2u_i}{dt^2} &=& (\nu t-u_i)+
l^2(u_{i+1}-2u_i+u_{i-1}) - (c+a\log(1+v_i)+b\log \theta_i)
\label{mme} \\
\frac{d\theta_i}{dt} &=& 1-v_i\theta_i
\end{eqnarray}
where $l \equiv (k_c/k_p)^{1/2}$ is 
the dimensionless stiffness parameter. 

With natural faults in mind, 
we give here rough estimates of various model parameters. 
Via the rise time of 
large earthquakes, the time unit $\omega ^{-1}$ may be estimated to be
a few seconds. Via the rupture-propagation speed,
the block size $d$ may be estimated to be a few kilometers 
\cite{Kawamura}.
The estimate of the characteristic slip
distance ${\cal L}$ remains to be largely ambiguous: Here,  we use  
an estimate by Scholz \cite{Scholzbook} and by Tse and Rice \cite{TseRice}, 
${\cal L}$ being of order a few mm or cm. 
Since the loading rate associated with 
the plate motion is typically a few 
cm/year, the dimensionless loading rate 
$\nu=\nu '/({\cal L}\omega)$ is of order $\nu \simeq 10^{-8}$.
The dimensionless quantity $k_p{\cal L}/{\cal N}$ may be written
in terms of the normal stress
$\sigma_n$, the density of the crust $\rho$ and other parameters defined
above as 
$\rho d\omega^2{\cal L}/\sigma_n$, 
which may be estimated
of order $10^{-4}$. 
The dimensionless parameter $c$ is then estimated
to be of order $10^3 \sim 10^4$. 
As mentioned, the parameters 
$a$ and $b$ are
one or two orders of magnitude smaller than $c$.
In the following, we set the parameter values of the model
around the typical values estimated above.

The coupled equations of motion (2) are solved numerically
by using the Runge-Kutta method of the fourth order. 
% The width of the time discretization $\Delta t$ is taken to be
% $\Delta t\nu =10^{-6}$. We have checked that the statistical properties 
% given below are unchanged
% even if we take the smaller $\Delta t$. 
Total number of $2\times 10^5$  events
are generated in each run,  which are used to perform various averagings,
while initial $10^4$ events are discarded as transients.
The total
number of blocks $N$ are taken to be $N=800$, 
with the open boundary condition. In some cases, larger systems with $N=1600$
are simulated to check the size dependence.
%The size dependence of the results was
%examined in Ref.\cite{MoriKawamura} with varying $N$ in the range
%$800\leq N \leq 6400$.
In the present simulation, we fix
the loading rate to $\nu =10^{-8}$.
The elastic parameter $l$ is set to $l=3$, which is the value
extensively studied by Mori and Kawamura 
for the 1D BK model with the Carlson-Langer velocity-weakening friction 
law \cite{MoriKawamura}.

%
%\begin{figure}[ht]
%\begin{figure}[htb]
%\begin{center}
%\includegraphics[scale=0.6]{magnitude-b.eps}
%\includegraphics[scale=0.6]{magnitude-a.eps}
%\includegraphics[scale=0.6]{magnitude-c.eps}
%\end{center}
%\baselineskip 3.2mm
%\caption{\small 
\begin{figure}
\onefigure[scale=0.7]{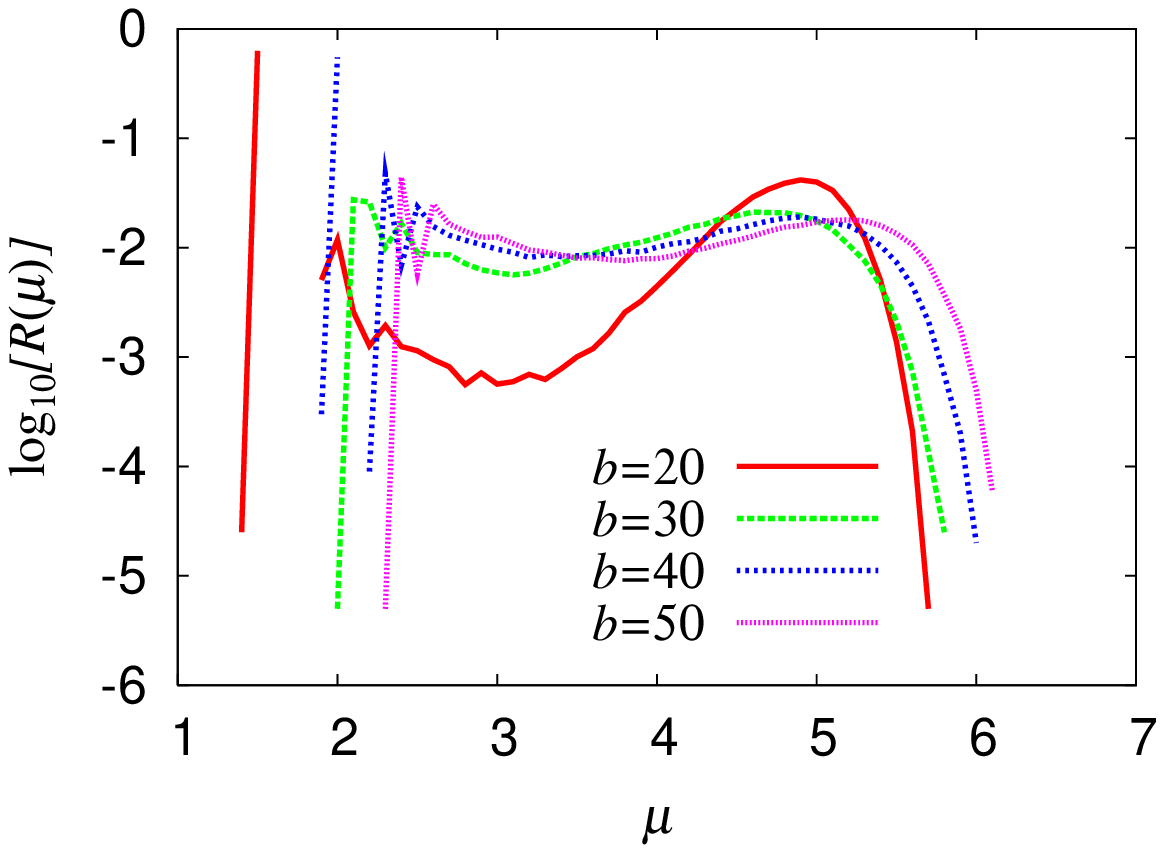}
\onefigure[scale=0.7]{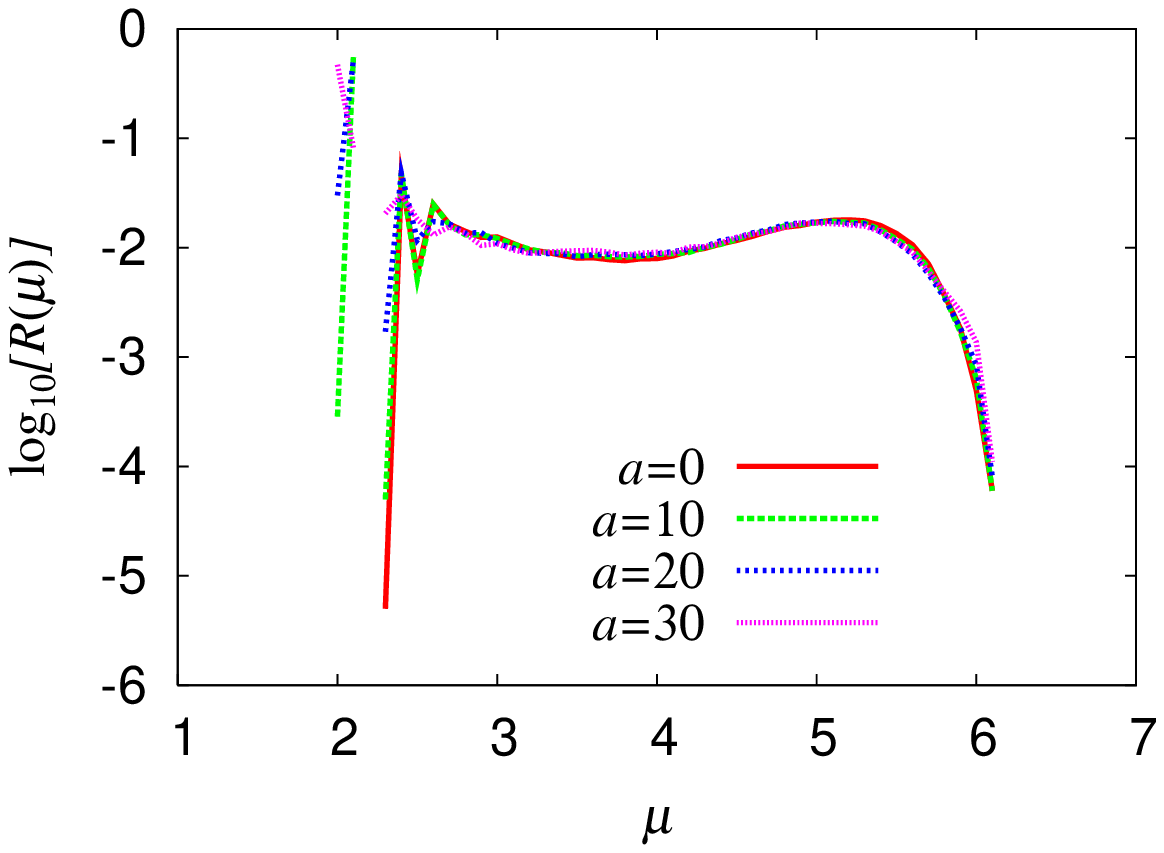}
\onefigure[scale=0.7]{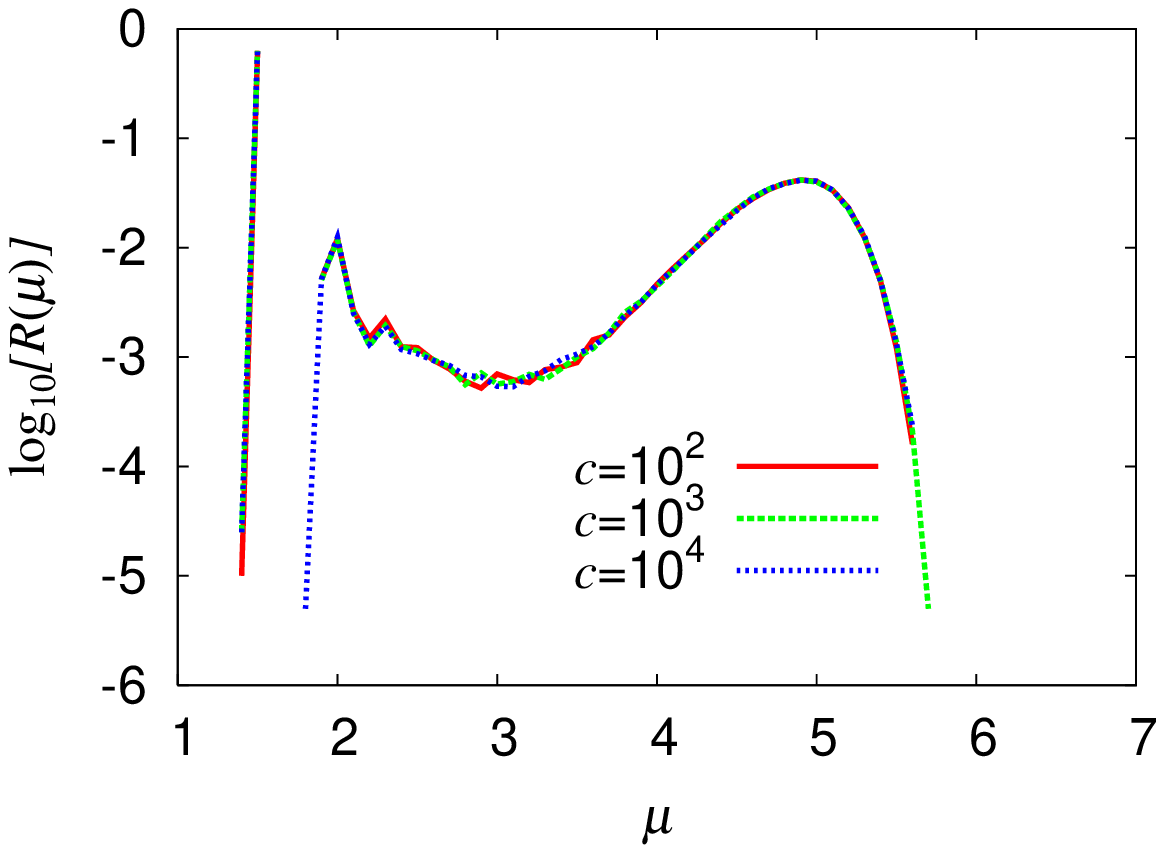}
\caption{
The magnitude distribution of seismic events.
In the upper figure,
the parameter $b$ is varied in the range $20\leq b\leq 50$ with fixing
$c=10^3$ and $a=0$.
In the middle figure,
the parameter $a$ is varied in the range $0\leq a\leq 30$ with fixing
$c=10^3$ and $b=50$. In the lower figure,
the parameter $c$ is varied in the range $10^2\leq c\leq 10^4$ with fixing
$a=0$ and $b=20$.
The other parameters are $\nu =10^{-8}$, $l=3$ and $N=800$ for
all cases. The magnitude distribution depends solely on $b$, not on $a$ and $c$.
}
%\label{fig1}
\end{figure}

In Figs.1(a)-(c), we show the magnitude distribution 
$R(\mu)$ of earthquakes
calculated from our numerical simulations. 
The magnitude of an event, $\mu$, 
is defined by 
$\mu=\log _{10} (\sum_i \Delta u_i)$, 
where $\Delta u_i$ is the 
total displacement of the $i$-th block during a given 
event and the sum is taken over all blocks involved in the event
\cite{Carlson91}.
%It should be noticed that
%the absolute numerics of the magnitude value of the
%BK model $\mu$ has no direct quantitative correspondence to the
%magnitude $m$ 
%of real earthquakes.
In all cases studied, the calculated $R(\mu)$ exhibits a continuous
nontrivial distribution from smaller to larger magnitudes.
%We note that such a continuous distribution for the size of earthquakes 
%are in
%apparent contrast to the result of Ref.[\textit{Tse and Rice}, 1986] 
%where only system-size
%large events were observed to occur quasi-periodically.

In the upper panel of Fig.1, 
the parameter $b$ is varied in the range $20\leq b\leq 50$
where the other constitutive parameters are fixed to $c=10^3$ and
$a=0$. As can be seen from the figure, the magnitude distribution 
depends on the $b$-value considerably. In the case of $b=20$, the distribution 
exhibits a pronounced peak at a large magnitude, with a pronounced
feature of a characteristic earthquake. In the case of $b=30\sim 50$, 
the distribution becomes flatter. In either case,
the distribution is not a power-law, largely deviating from the Gutenberg-Richter (GR) law.
We also made a preliminary study of the smaller $b$ region ($b<20$) and of
the larger $b$ region ($b>50$). 
In the case of smaller $b$, an event tends to occur
as a creep-like small event rather than a stick-slip, whereas, 
in the case of larger $b$, more weight is given 
to system-size large events in $R(\mu)$ and the data suffer from 
significant finite-size effects.

In the middle panel of Fig.1, 
the parameter $a$ is varied in the range $0\leq a\leq 30$
with $c=10^3$ and $b=50$.
The magnitude distribution hardly depends 
on the $a$-value, indicating that the relavant parameter here is 
$b$ itself, rather than $a-b$ as widely believed \cite{Scholzbook,ScholzRev}. 
In the lower panel of Fig.1, 
the parameter $c$ is varied in the range $10^2\leq c\leq 10^4$
with $a=0$ and $b=20$.
Again, the magnitude distribution 
hardly depends on the $c$-value. 
%This suggests that the properties of 
%stick-slip dynamics is primarily determined by the rate- and state-dependent 
%part of the friction force, even though the total magnitude of the friction 
%is dominated by the constant term $c$. 
Hence, it turns out that the magnitude distribution
is sensitive only to the $b$-value. 
%Intuitively, 
%this is easy to understand since the frictional instability  
%in the present model is borne
%by the $b$-term.

We also calculate the local recurrence-time distribution of large events 
in order to see how large events 
repeat in time: In defining the recurrence time locally,
the subsequent large event is counted when a large
event occurs with its epicenter in the region within
10 blocks from the epicenter of the previous large event.
%The picture of a characteristic
%earthquake presumes the existence of a characteristic recurrence time. 
%In this case, the distribution of the recurrence time of large earthquakes, 
%$T$, 
%is expected to exhibit a peak structure at such a characteristic time scale.
%If the SOC concept applies to large earthquakes, by contrast,
%such a peak structure would not show up. 
In Figs.2, we show the distribution of the
local recurrence time $T$ of large earthquakes whose magnitude is greater
than $\mu \geq \mu _c=4$ for the case of $b=20$ (upper panel) and  $b=40$ 
(lower panel). The other constitutive
parameters are fixed to $c=10^3$ and $a=0$.

As can be seen from the upper panel of Fig.2, 
the distribution in the case of $b=20$
shows a sharp peak at $T\nu \simeq 500$, indicating a pronounced periodic
feature of earthquake recurrence. Smaller peaks are also discernible
at multiples of an elementary period. The inset represents a magnified
view of the sharp peak at $T\nu \simeq 500$, where the data both for 
$N=800$ and $1600$ are shown. 
The peak tends to get lower with $N$. In order
to evaluate the possible boundary effect, we also calculate the recurrence-time
distribution only for the subset of events which stay
in the interior of the
system, {\it i.e.\/}, the events whose rupture does not reach the boundary,
and the results are 
shown in the inset for both $N=800$ and 1600. The data for these ``interior
events''  no longer show appreciable size effect
suggesting that this sharp peak might persist even in an infinite-size limit. 

Somewhat different behavior is observed in the case of $b=40$.
As can be seen from the lower panel of Fig.2, the distribution
exhibits {\it two\/} independent peaks
at $T\nu \simeq 400$ and at $T\nu \simeq 1000$.
% , with some multiples of them.
Essentially the same behavior is also observed for $b=30$ and 50.
The inset represents a magnified
view of the second peak at $T\nu \simeq 1000$.
While the peak tends to get 
lower with increasing $N$, it is still clearly visible even at $N=1600$.
% persists even in an infinite-size limit as in the case of $b=20$. 
The first peak located at $T\nu \simeq 400$, by contrast, 
does not show appreciable size effect. 
%Hence, the existence of two independent characteristic periods 
%seems to be a property of an infinite system. 
The observed double-peak structure means that events 
tend to occur with double periods, although
the origin of such a doubly-periodic behavior is not clear at the 
present stage.

%\begin{figure}[htb]
%\begin{center}
%\includegraphics[scale=0.65]{recurrence-b20.eps}
%\includegraphics[scale=0.65]{recurrence-b40.eps}
%\end{center}
%\baselineskip 3.2mm
\begin{figure}
\onefigure[scale=0.7]{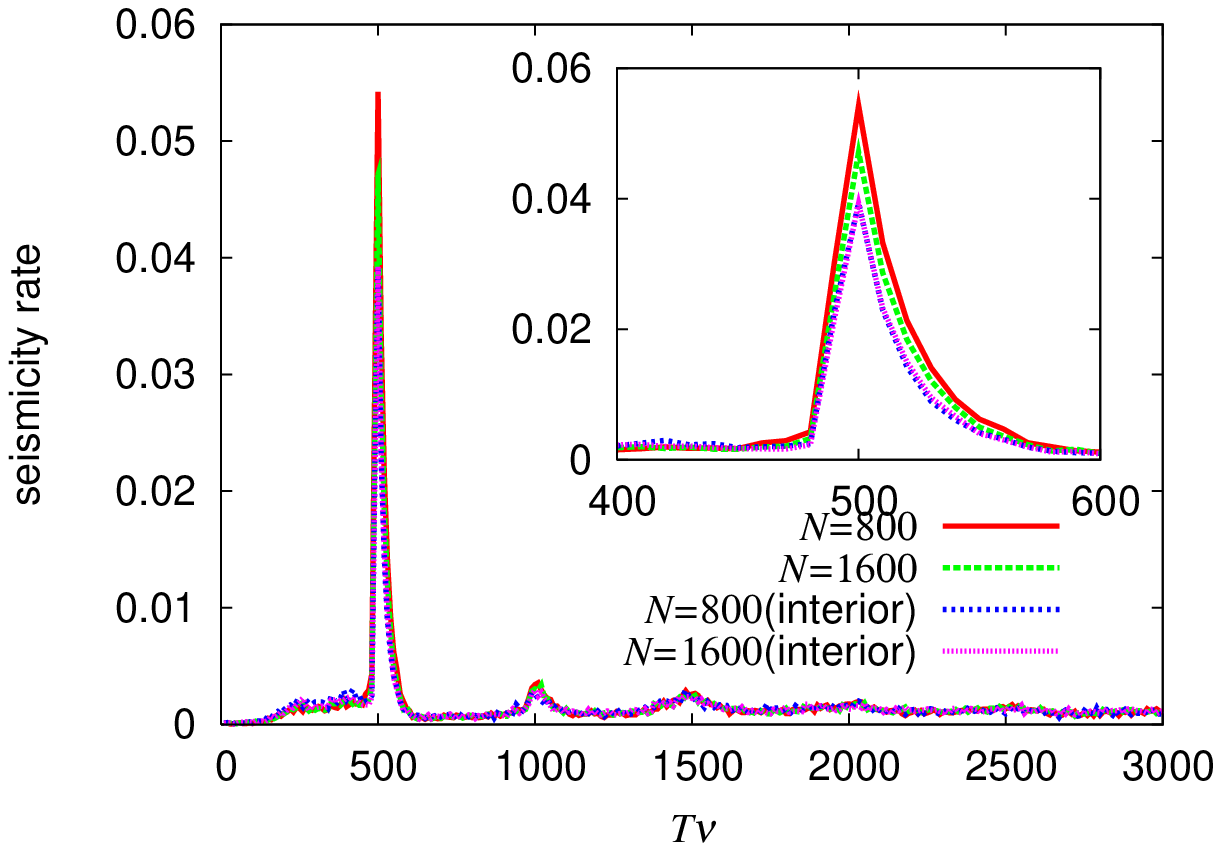}
\onefigure[scale=0.7]{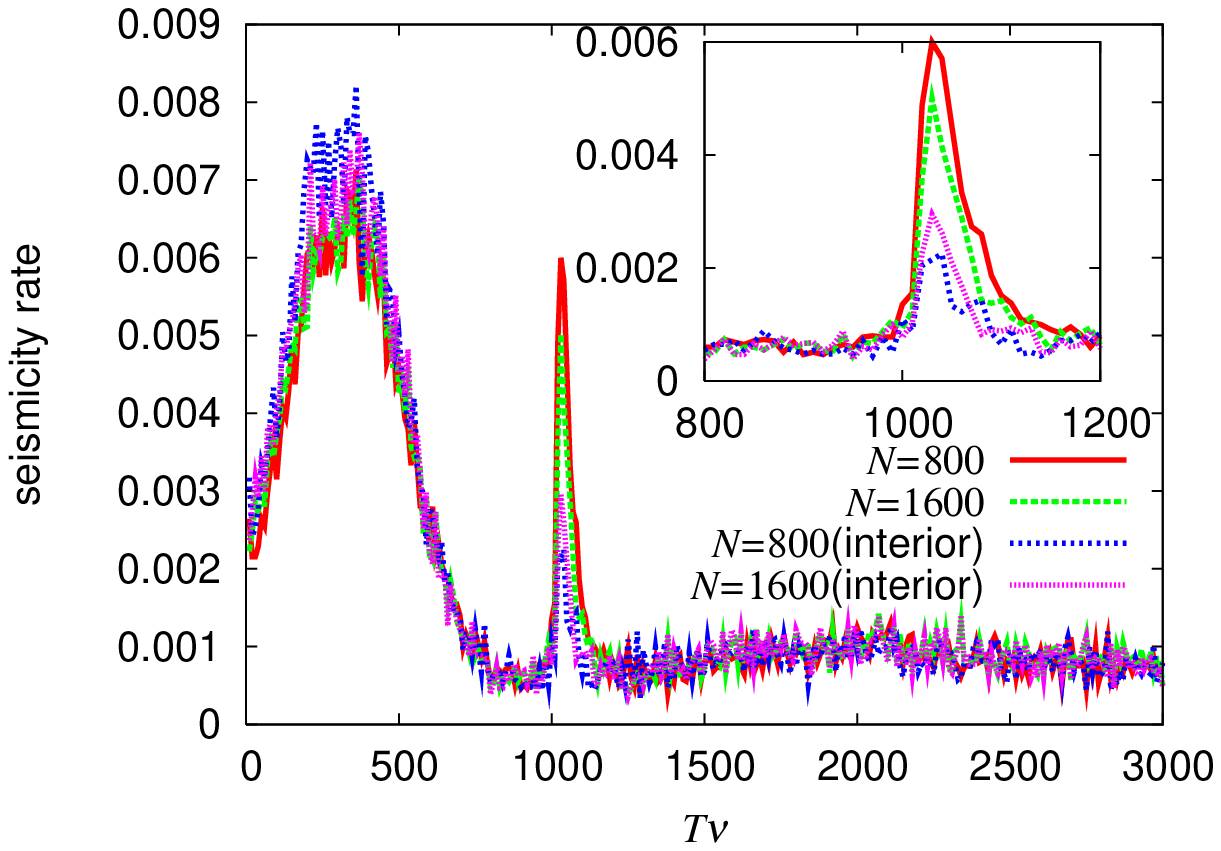}
%\caption{\small 
\caption{
The local recurrence-time distribution of large events of $\mu \geq
\mu_c=4$ for the case of $b=20$ (upper panel) and of $b=40$ (lower panel). 
The other parameters are $c=10^3$, $a=0$, $\nu=10^{-8}$ and $l=3$.
The system size is $N=800$ or $1600$.
The insets represent magnified views of the peak region. 
The data represented as ``interior'' are the ones taken for the 
type of large events 
whose rupture does not reach the boundary of the system.  The observed distinct peak structure suggests near periodic occurrence of events. 
}
%\label{fig2}
\end{figure}

In summary, 
we studied the statistical properties
of the 1D BK model combined with a rate- and state-dependent 
constitutive law. 
We found the followings: (i) A continuous size distribution, 
spanning from smaller
to larger magnitudes, has been obtained. Somewhat unexpectedly, however, the power-law distribution like the GR law 
is not realized in this model, in spite of the fact that the rate- and
state-dependent friction law employed here is the standard one widely
used in recent earthquake studies. 
(ii) Among several parameters characterizing the 
rate- and state-dependent constitutive law, the statistical properties of
earthquakes  depend most sensitively on the parameter $b$, which
describes the extent of the state-dependent frictional instability.
(iii) The recurrence-time distribution indicates the near
periodic recurrence of large events. 
%This is consistent with an earlier observation 
%by Tse and Rice \cite{TseRice} and by Rice \cite{Rice}.
For the case of stronger frictional instability,
doubly-periodic recurrence characterized by the two
independent periods becomes eminent.

These observations suggest that earthquakes described by the the present model 
possesses pronounced ``characteristic'' features.
The characteristic tendency of our present model
is even more enhanced than the one observed for the BK model
with the Carlson-Langer velocity-weakening constitutive law.
Such pronounced characteristic features are in apparent
contrast to the observations for natural 
faults, since the analysis of real seismic
catalog is known to lead to
more ``critical'' behaviors, {\it e.g.\/}, 
the GR law for the magnitude distribution: See, {\it e.g.\/}, 
Ref.\cite{Kawamura}.

Identifying the cause of this deviation would be of
particular importance. First, we note that
the rate- and state-dependent constitutive
law was derived based on the data of rock friction 
in laboratory measurements. 
One has to be careful that the relevant length scale in laboratory 
measurements differs from 
the one at natural faults by orders of magnitude. 
There is no obvious guarantee
that the rate- and state-dependent constitutive
law derived from laberatory experiments is also valid in describing 
natural faults. 
Other possible reason of the deviation
may be an intrinsic discreteness of the present model. According to
Refs.\cite{Rice,Ben-ZionRice},
however, the discreteness tends to 
enhance the criticality, which is just opposite to what is
required to account for the observed deviation.
We also need to
recognize that real earthquake catalog is usually taken not for a single
fault, but over many faults. There has been a suspicion that, even if
the property of a single uniform fault is characteristic, 
the property obtained after averaging over many faults, each of which
has different characteristics, might become apparently critical.
If this is really the case,  the observations for natural faults 
is not necessarily inconsistent with the observations
for the present model, since the BK model deals with the property of
a single uniform fault. 
In any case, further studies are required to settle the issue.


\begin{thebibliography}{99}

\bibitem{Scholzbook}  
\Name{Scholz C.H.}
\Book{The Mechanics of Earthquakes and Faulting} 
\Publ{Cambridge Univ. Press}
\Year{1990}

\bibitem{ScholzRev}  
\Name{Scholz C.H.} 
\REVIEW{Nature}{391}{1998}{3411}.

\bibitem{Dieterich} 
\Name{J.H. Dieterich} 
\REVIEW{J. Geophys. Res.}{77}{1972}{3690};
\REVIEW{J. Geophys. Res.}{84}{1979}{2161}.

\bibitem{Ruina}  
\Name{A. Ruina} 
\REVIEW{J. Geophys. Res.}{88}{1983}{10359}.

\bibitem{BK}  
\Name{R. Burridge \and L. Knopoff}
\REVIEW{Bull. Seismol. Soc. Am.}{57}{1967}{3411}.

\bibitem{CL} 
\Name{Carlson J.M. \and Langer J.S.} 
\REVIEW{Phys. Rev. Lett.}{62}{1989}{2632}; 
\REVIEW{Phys. Rev. A}{40}{1989}{6470}.

\bibitem{Carlson91} 
\Name{Carlson J.M., Langer J.S., Shaw B.E. \and Tang C.}
\REVIEW{Phys. Rev. A}{44}{1991}{884}.

\bibitem{Carlson94} 
\Name{Carlson J.M., Langer J.S. \and Shaw B.E.}
\REVIEW{Rev. Mod. Phys.}{66}{1994}{657}.

\bibitem{Shaw} 
\Name{Shaw B.E.}
\REVIEW{J. Geophys. Res.}{100}{1995}{18239}.

\bibitem{Myers} 
\Name{Myers C.R., Shaw B.E. \and Langer J.S.}
\REVIEW{Phys. Rev. Lett.}{77}{1996}{972}.

\bibitem{MoriKawamura} 
\Name{Mori T. \and Kawamura H.} 
\REVIEW{Phys. Rev. Lett.}{94}{2005}{058501}; 
\REVIEW{J. Geophys. Res.}{111}{2006}{B07302}.

\bibitem{Kawamura} 
\Name{Kawamura H.} 
\Book{Modelling critical and catastrophic phenomena in geoscience}
\Editor{P. Bhattacharyya \and B. Chakrabarti}
\Publ{Springer}
\Year{2006}
\Page{223}
(cond-mat/0603335). 

\bibitem{Xia} 
\Name{Xia J., Gould H., Klein W. \and  Rundle J.B.} 
\REVIEW{Phys. Rev. Lett.}{95}{2005}{248501}; cond-mat/0601679.

\bibitem{Ida} 
\Name{Ida Y.}
\REVIEW{J. Geophys. Res.}{77}{1972}{3796}.

\bibitem{TseRice} 
\Name{Tse S.T. \and Rice J.R.} 
\REVIEW{J. Geophys. Res.}{91}{1986}{9452}.

\bibitem{Stuart} 
\Name{Stuart W.D.} 
\REVIEW{Pure. Appl. Geophys.}{126}{1988}{619}.

\bibitem{Horowitz} 
\Name{Horowitz F. \and Ruina A.} 
\REVIEW{J. Geophys. Res.}{94}{1989}{10279}.

\bibitem{Rice} 
\Name{Rice J.R.}
\REVIEW{J. Geophys. Res.} {98}{1993}{9885}.

\bibitem{Ben-ZionRice} 
\Name{Ben-Zion Y. \and Rice J.R.}
\REVIEW{J. Geophys. Res.}{102}{1997}{17771}.

%\bibitem{KotoHirasawa} N. Kato and T. Hirasawa,
%Bull. Seismol. Soc. Am. {\bf 89}, 1401-1417 (1999).

\bibitem{Kato} 
\Name{Kato N.}
\REVIEW{J. Geophys. Res.}{109}{2004}{B12306}.

\bibitem{Bizzarri} 
\Name{Bizzarri A. \and Cocco M.} 
\REVIEW{J. Geophys. Res.}{111}{2006}{B05303}; {B05304}.

\bibitem{Cochard} 
\Name{Cochard A. \and Madariaga R.} 
\REVIEW{J. Geophys. Res.} {101}{1996}{25321}.

\bibitem{CaoAki}  \Name{Cao T. \and  Aki K.} \REVIEW{Pure. Appl. Geophys.} {124}{1986}{487}.



\end{thebibliography}
\end{document}